\documentclass[conference]{IEEEtran}
\usepackage{cite}

\ifCLASSINFOpdf
  \usepackage[pdftex]{graphicx}
\else
  \usepackage[dvips]{graphicx}
\fi
\usepackage{amsmath}
\usepackage{bbm}

\usepackage{bm}

\IEEEoverridecommandlockouts

\usepackage{tikz}

\newcommand\copyrighttext{%
	\footnotesize \textcopyright 2019 IEEE. Personal use of this material is permitted.
	Permission from IEEE must be obtained for all other uses, in any current or future
	media, including reprinting/republishing this material for advertising or promotional
	purposes, creating new collective works, for resale or redistribution to servers or
	lists, or reuse of any copyrighted component of this work in other works.}
\newcommand\copyrightnotice{%
	\begin{tikzpicture}[remember picture,overlay]
	\node[anchor=south,yshift=10pt] at (current page.south) {\fbox{\parbox{\dimexpr\textwidth-\fboxsep-\fboxrule\relax}{\copyrighttext}}};
	\end{tikzpicture}%
}

\begin{document}
%
\title{Conservative Link Adaptation for Ultra Reliable Low Latency Communications \thanks{The research has been carried out at IITP RAS and supported by the Russian Government (Contract No 14.W03.31.0019) and Huawei Technologies}}

\author{\IEEEauthorblockN{
	Andrey Belogaev\IEEEauthorrefmark{1},
	Evgeny Khorov\IEEEauthorrefmark{1},
	Artem Krasilov\IEEEauthorrefmark{1},
	Dmitri Shmelkin\IEEEauthorrefmark{2}, and
	Suwen Tang\IEEEauthorrefmark{2}
}
\IEEEauthorblockA{\IEEEauthorrefmark{1}Institute for Information Transmission Problems, Russian Academy of Sciences, Moscow, Russia\\
Email: \{belogaev, khorov, krasilov\}@iitp.ru}
\IEEEauthorblockA{\IEEEauthorrefmark{2}Huawei Technologies Moscow Research Center, Moscow, Russia\\
Email: \{shmelkin.dmitri, tangsuwen\}@huawei.com}
}


%


\maketitle

\copyrightnotice

\begin{abstract}
	Ultra reliable low latency communications (URLLC) is one of the most promising and demanding services in 5G systems. This service requires very low latency of less than $1-10$ ms and very high transmission reliability: the acceptable packet loss ratio is about $10^{-5}$. To satisfy such strict requirements, many issues shall be solved. This paper focuses on the link adaptation problem, i.e., the selection of a modulation and coding scheme (MCS) for transmission based on the received channel quality indicator (CQI) reports. On the one hand, link adaptation should select a robust MCS  to provide high reliability. On the other hand, it should select the highest possible MCS to reduce channel resource consumption. The paper shows that even for one URLLC user, link adaptation is still a challenging problem, especially in highly-variant channels. To solve this problem, a conservative link adaptation algorithm is designed. The algorithm estimates the strongest channel degradation at the time moment of the actual packet transmission and selects an MCS taking into account the worst degradation. The obtained results show that the proposed algorithm is efficient in terms of both the packet loss ratio and the channel resource consumption.  
\end{abstract}


%
\IEEEpeerreviewmaketitle

\vspace{-0.2cm}
\section{Introduction}
\vspace{-0.1cm}

It is well-known that the next generation of wireless networks (5G) is expected to support a wide range of new services. Ultra reliable low latency communications (URLLC) is an example of such services that implies very strict requirements on packet delivery delay ($1-10$ ms) and packet loss ratio ($PLR<10^{-5}$)~\cite{itu_urllc}. 

To provide such a low PLR, the link adaptation algorithm (i.e., the algorithm which selects modulation and coding scheme (MCS) according to the current link quality) should be robust to any change of channel conditions. In LTE and 5G systems, link adaptation works as follows. User Equipment (UE) measures the signal-to-noise ratio (SNR), maps these measurements to channel quality indicators (CQIs) and then reports them to a base station. Using the received CQI reports, the base station (gNB) selects such an MCS, which provides block error rate (BLER) less than some target value (in LTE it is set to 10\%). Apparently, the latest CQI report cannot give an accurate estimation of SNR, and many works in the literature focus on robust link adaptation schemes correcting such inaccuracies. Moreover, the number of CQI values is smaller than the number of available MCSs. 

To address these problems, a popular algorithm of the outer loop link adaptation (OLLA)~\cite{olla, olla_optimized} uses hybrid automatic repeat request (HARQ) statistics on positive and negative acknowledgments (ACK/NACK) to adjust MCS at the base station. Specifically, OLLA adds some offset to SNR estimation, which increases on each received ACK and decreases on each NACK. However, slow convergence of this approach makes it inapplicable to URLLC scenarios.

In~\cite{pedersen_urllc}, the authors consider link adaptation for URLLC traffic in the presence of interference and do not pay much attention to the estimation of fading. However, our results show that even for slowly moving UEs in the scenario without interference, we have to take into account fading when predicting channel degradation in order to provide high reliability.

The authors of~\cite{pedersen_urllc_with_embb} propose a scheduling algorithm with dynamic BLER adjustment for URLLC and eMBB (enhanced mobile broadband) traffic. In particular, they divide the scheduling process into two steps: (i)  scheduling of the minimal amount of channel resource, which is enough to transmit URLLC data with a given target BLER; (ii) scheduling of the remaining channel resource in order to transmit data with a more conservative (and thus more reliable) MCS. In this paper, we use a similar idea of using a more conservative MCS than the latest CQI report suggests. However, in contrast to~\cite{pedersen_urllc_with_embb}, we take into account the available URLLC packet delay budget.  

In LTE, the delay increases with the number of performed HARQ retransmissions since each retransmission is sent only after receiving NACK or after some timeout. In~\cite{learning_feedbackless}, the authors propose to use the so-called \emph{feedbackless} transmissions for URLLC. Specifically, before any transmission, the base station chooses one of the two options: (i) normal feedback-based transmission allowing $N$ transmission attempts (initial transmission and $N-1$ HARQ retransmissions); (ii) one-shot transmission with $N$ times lower code rate. For that, the authors of the paper use a reinforcement learning approach that aims at minimizing the expected long-term cost, which depends on the latency and the reliability of transmission. However, the authors assume that the code rate is constant. In contrast, we focus on link adaptation and an appropriate MCS selection.

The paper \cite{two_blers} studies dynamic adjustment of the target BLER for initial transmission and a HARQ retransmission of a URLLC packet. Specifically, the authors propose to send two CQI reports corresponding to two different target BLERs. These target BLERs should be selected in such a way to make the initial transmission efficient and to provide the required reliability with HARQ retransmission on more robust MCS. Despite the benefits achieved with this approach, there are still some challenges in its applicability to URLLC since it strongly relies on accurate channel estimation. 

A recent 3GPP report from Nokia~\cite{nokia_report} states that in a highly-variant channel, it is beneficial to know the worst case SNR conditions experienced by the UE at a given time. Unfortunately, the specific algorithm for MCS selection is considered for future study.  Inspired by their idea, we propose a conservative link adaptation algorithm, which estimates the maximal channel degradation based on the difference between CQI reports obtained at the gNB and shows good performance in terms of PLR, latency and channel resource consumption.

The rest of the paper is organized as follows. In Section~\ref{sec:algo}, we describe the proposed algorithm. We evaluate the performance of this algorithm and analyze the obtained results in Section~\ref{sec:results}. Section~\ref{sec:conclusion} concludes the paper.

\section{Proposed Algorithm}
\label{sec:algo}

\subsection{Main Idea}

\begin{figure}[!t]
	\centering    
	\includegraphics[width=0.9\columnwidth]{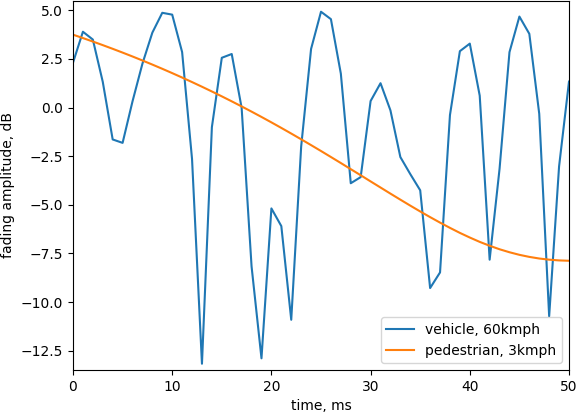}
	\vspace{-0.3cm}
	\caption{\label{fig:fadings}Rayleigh fading.}
	\vspace{-0.5cm}
\end{figure}

\vspace{-0.1cm}
Fig.~\ref{fig:fadings} illustrates how the channel quality of one resource block (RB) varies with time. We can see that for high speeds of UEs, the channel quality changes very fast. This complicates the problem of MCS selection since the optimal MCS shall (i) provide high success probability, and (ii) be as high as possible to consume channel resource frugally. However, in URLLC scenarios, we have a very low delay budget for transmission, which is enough only for one or two transmission attempts for each packet. In this case, we cannot rely solely on the last report from the UE because it can be non-relevant at the time of actual transmission. 

To address this problem, we propose a conservative approach which is based on the estimation of the worst-case conditions. In particular, we estimate the maximal channel degradation during the interval between the instant when the CQI was measured and the instant, when the transmission was performed. Below we describe the algorithm in detail.

\subsection{CQI estimation}

The UEs divide the whole band into several subbands containing the integer number of RBs. Once in a CQI reporting period $T_{CQI}$ the UEs calculate the average SNRs in each subband, map obtained SNRs to integer CQIs and report them to the gNB.

\begin{figure}[!t]
	\centering    
	\includegraphics[width=0.9\columnwidth]{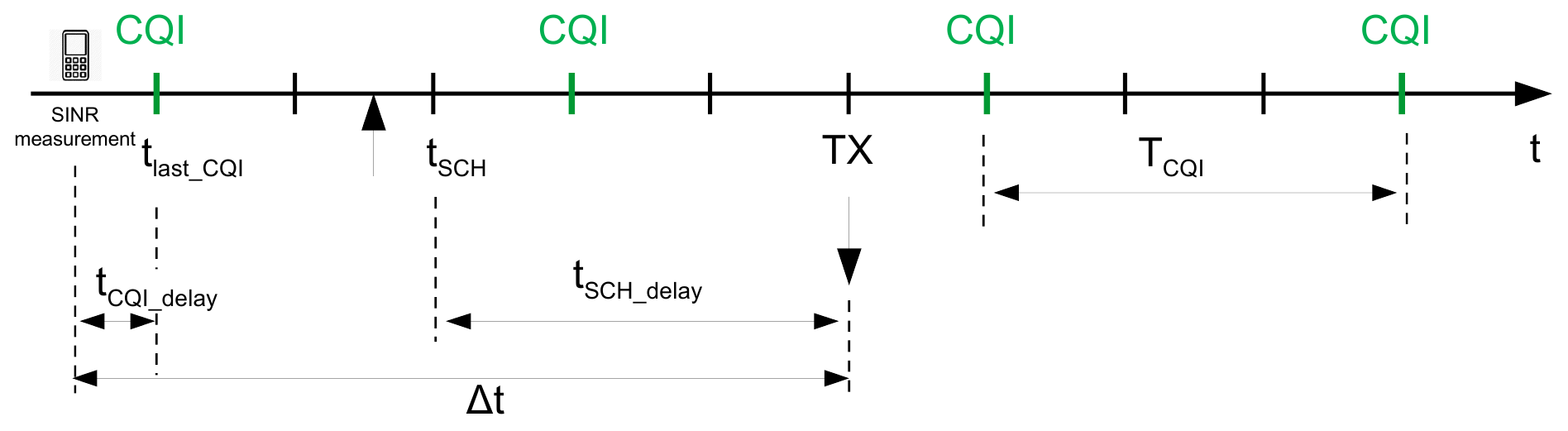}
	\vspace{-0.3cm}
	\caption{\label{fig:denotions}CQI estimation.}
	\vspace{-0.5cm}
\end{figure}

The amount of data which can be successfully delivered in an RB depends on its current state. During the scheduling procedure at time moment $t_{SCH}$, the channel quality is estimated using previously received CQI reports. Let $t_{last\_CQI}$ be the time moment when the last CQI report has been received by the gNB. Then, when actual data transmission is performed, the last received CQI is outdated by $\Delta t = t_{SCH} + t_{sch\_delay} - t_{last\_CQI} + t_{CQI\_delay}$, where $t_{sch\_delay}$ is scheduling delay which is needed for scheduling procedure and preparing data for transmission and $t_{CQI\_delay}$ is the delay needed for CQI report generation and transmission (see Fig.~\ref{fig:denotions}). 

Since the channel quality can significantly change with time, the estimation error which we obtain using such an outdated CQI report can be very high. Hence, we propose to estimate channel quality conservatively assuming that the channel degradation between time moments $t$ and $t - \Delta t$ is less than the maximal channel degradation observed between time moments $t'$ and $t' - \Delta t$ for any $t' \in (t-W, t)$, where $W$ is the observation window, i.e., 

\vspace{-0.5cm}
\begin{align*}
\nonumber \Delta CQI (\Delta t) &= \max \left[ CQI(t' - \Delta t) - CQI (t') \right] \\ \nonumber &\forall\; t': t-W < t' < t
\end{align*}
\vspace{-0.5cm}

Specifically, we estimate CQI in each subband as follows:

\vspace{-0.5cm}
\begin{align}
\label{eq:cqi_diff}
CQI (t_{SCH}) = \max (0, CQI (t_{last\_CQI}) - \Delta CQI (\Delta t))
\end{align}       
\vspace{-0.5cm}

To calculate $\Delta CQI (\Delta t)$, we collect statistics on previously received CQI reports. Obviously, the maximal value of $\Delta t$ equals $\Delta t_{max} = t_{sch\_delay} + T_{CQI} + t_{CQI\_delay}$, when the CQI report is obtained right after scheduling procedure at time moment $t_{SCH}$. Therefore, to estimate the channel degradation during any $\Delta t$ on each received CQI report, we need to collect statistics $CQI\_HISTORY$ only on last $\left\lceil \frac{\Delta t_{max}}{T_{CQI}} \right\rceil$ CQI reports. Then, on each received CQI report $last\_CQI$, we calculate $\Delta CQI = CQI\_HISTORY[t_{last\_CQI} - \Delta t] - last\_CQI$ and replace in $CQI\_HISTORY$ CQI value corresponding to time moment $t_{last\_CQI} - \Delta t_{max}$ with $last\_CQI$.

In order to estimate the maximal channel degradation, we calculate the maximum value of $\Delta CQI$ in a sliding window of size $W/T_{CQI}$. In other words, for each $\Delta t$ (quantized with step $T_{CQI}$) we keep the maximal value of the observed channel quality degradation during the time interval $W$. 

Using the described above procedure, we can estimate $\Delta CQI$ for each subband separately. In scenarios, where noise and/or interference affects the whole band, i.e., all subbands in the same manner, we can merge statistics collected for all subbands and find the maximal channel degradation faster (using a lower value of $W$). For that, we take the maximum value from $\Delta CQI$ values measured in different subbands.

\subsection{MCS Selection and Resource Allocation}

When scheduling algorithm makes a decision on allocating RBs to UEs, it should take into account the channel conditions in a certain subband corresponding to the moment of actual transmission, i.e., utilize CQI values calculated according to~\eqref{eq:cqi_diff}. However, the result of the subtraction is truncated by zero, so that we lose information about the RBs with reported CQIs close to zero. Hence, in the scheduling algorithm we use both the last reported CQIs and the estimated CQIs as follows.

\begin{enumerate}
	\item For each RB, calculate the scheduling metric for each UE (e.g., with LLC-PF scheduler designed in~\cite{WCNC_URRLC} for URLLC) and sort UEs according to this metric.
	\item For each RB, remove from consideration those UEs that have reported CQI 0 (according to last reported CQIs) for this RB. 
	\item Find leaders for each RB, i.e., UEs with the highest scheduling metric.
	\item Sort RBs in the descending order of CQIs reported by corresponding leaders.
	\item Starting from the first RB (with the highest CQI value), allocate RBs to the corresponding leader.
	\item For each UE, calculate the maximal transport block (TB) size, i.e., the number of bits which can be transmitted with allocated to this user RBs, using CQI estimation (as described below). All RBs assigned to the UE at step (5) and not used are considered for further scheduling.
	\item Remove from consideration the UEs that can transmit all buffered data in already allocated RBs. 
	\item Go to step (3) and continue procedure until all RBs are allocated or none of the remaining RBs can increase the TB size for any UE. 
\end{enumerate}

Let us describe the maximal TB size search procedure at step (6). Consider  a UE with the allocated set of RBs $\{a_1, a_2, ..., a_n\}$. 

\begin{enumerate}
	\item For each RB $a_i$ from the set, we calculate the CQI estimation according to equation~\eqref{eq:cqi_diff} and map the obtained CQIs $\{CQI^1, ..., CQI^n\}$ to Signal-to-noise-ratios $\{SNR^1, ..., SNR^n\}$. Below we assume that RBs are sorted in the descending order of SNRs.
	\item We calculate a single effective $SNR_k$ (see~\cite{effective_SNR} for details) for each subset $A_{k} = \{a_1, ..., a_k\}, k = \{1,...,n\}$.
	\item Using the effective $SNR_k$, we find such an MCS $MCS_{k}$ that  allows obtaining BLER less than the given target value.
	\item Assuming that $MCS_k$ is used in all RBs, we calculate the TB size $TB_{k}$ for each subset $A_{k}$.
	\item We find the subset $A_{k}$ providing the maximal TB size. 
\end{enumerate}

Since URLLC traffic implies strict delay and PLR requirements, to avoid packet losses, we do not use RBs with reported CQI 0 in the described above procedure. However, for the UE that cannot deliver packet before its deadline and has selected the lowest MCS (i.e., MCS 0) for transmission these RBs can be useful. Hence, after the end of the described above procedure we do the following steps.

\begin{enumerate}
	\item For each UE, we check whether it can deliver its packets before their deadlines.
	\item For those UEs which cannot transmit a packet before deadline, we calculate the TB size which is needed to meet the deadline and check whether we can obtain such TB size using MCS 0 and RBs having CQI equal 0.
	\item If the required TB size can be obtained, we select MCS 0 and add to previously selected RBs the minimal number of RBs which allow satisfying the UE demands.
\end{enumerate}
   
\section{Numerical Results}
\label{sec:results}

\subsection{Simulation Setup}
\label{subsec:setup}

To evaluate the performance of the proposed link adaptation algorithm, we use a well-known discrete-event simulator NS-3~\cite{ns-3} with implemented features enabling URLLC (see~\cite{WCNC_URRLC}). Simulation parameters are summarized in Table~\ref{tab:sim-params}. Similar to~\cite{WCNC_URRLC}: 

\begin{itemize}
	\item We consider a mini-slots time structure with the mini-slot duration of 2 OFDM symbols, where 25\% of mini-slot is used for control data and reference signals and 75\% is used for user data.
	\item The time interval between the initial transmission of a packet and its HARQ retransmission equals 3 mini-slots.
	\item For each URLLC packet, we consider the delay budget of 1 ms (i.e., seven mini-slots). Therefore, only one HARQ retransmission can be performed within the given delay budget. 
	\item The control channel is reliable, i.e., HARQ feedback and CQI report messages are always successfully delivered.  
\end{itemize}  

In experiments, we consider a single gNB and a single UE. The UEs receives  URLLC traffic in downlink with packet inter-arrival time of 3 ms. Obviously, packet transmission latency and reliability depend on the channel quality. Hence, we vary the Geometry factor, which equals SNR experienced by a UE in the same position without fading. For that, we vary the distance between the UE and the gNB.    

\begin{table}[t]
	\caption{Simulation parameters} \label{tab:sim-params}
	\centering
	\begin{tabular}{|l|l|}
		\hline    \textbf{Parameter} & \textbf{Value} \\  
		\hline
		Carrier frequency & 2 GHz \\
		Bandwidth & 20 MHz \\
		Subcarriers spacing & 15 kHz \\
		RB width & 12 subcarriers \\
		OFDM symbol duration & $71.4$~us \\ 
		Mini-slot duration & 2 OFDM symbols \\
		\parbox{3.3cm}{Number of HARQ retransmissions per packet\\} & 1 \\   
		gNB/UE TX power & 30 / 23 dBm  \\
		gNB antenna pattern & omnidirectional \\
		\parbox{3.3cm}{Fading model \\} &  \parbox{4.2cm}{Rayleigh multipath fading (Extended Pedestrian/Vehicle A)\\} \\
		Geometry factor & -3..25 dB \\
		\hline
	\end{tabular} 
	\vspace{-0.5cm}
\end{table} 

We compare the following solutions:

\begin{enumerate}
	\item \emph{last CQI} estimation, which uses only last reported CQIs for MCS selection;
	\item \emph{conservative} link adaptation algorithm described in Section~\ref{sec:algo} with $W/T_{CQI} = \{10, 100\}$;
	\item \emph{MCS0 in best RBs} which always selects MCS 0 and allocates the best RBs such that the obtained TB size is greater than the current queue size, or allocates all RBs if it is impossible.
\end{enumerate}

For performance evaluation, we consider the following key performance indicators.

\begin{enumerate}
	\item PLR is the fraction of packets which delivered after the deadline or have been lost;
	\item The average MCS used for transmissions.
	\item RB usage is the percentage of RBs used for data transmission. It characterizes  channel resource consumption.
\end{enumerate}

\subsection{Analysis of Results}
\label{subsec:analysis}

\begin{figure}[!t]
		\center{\includegraphics[width=0.7\linewidth]{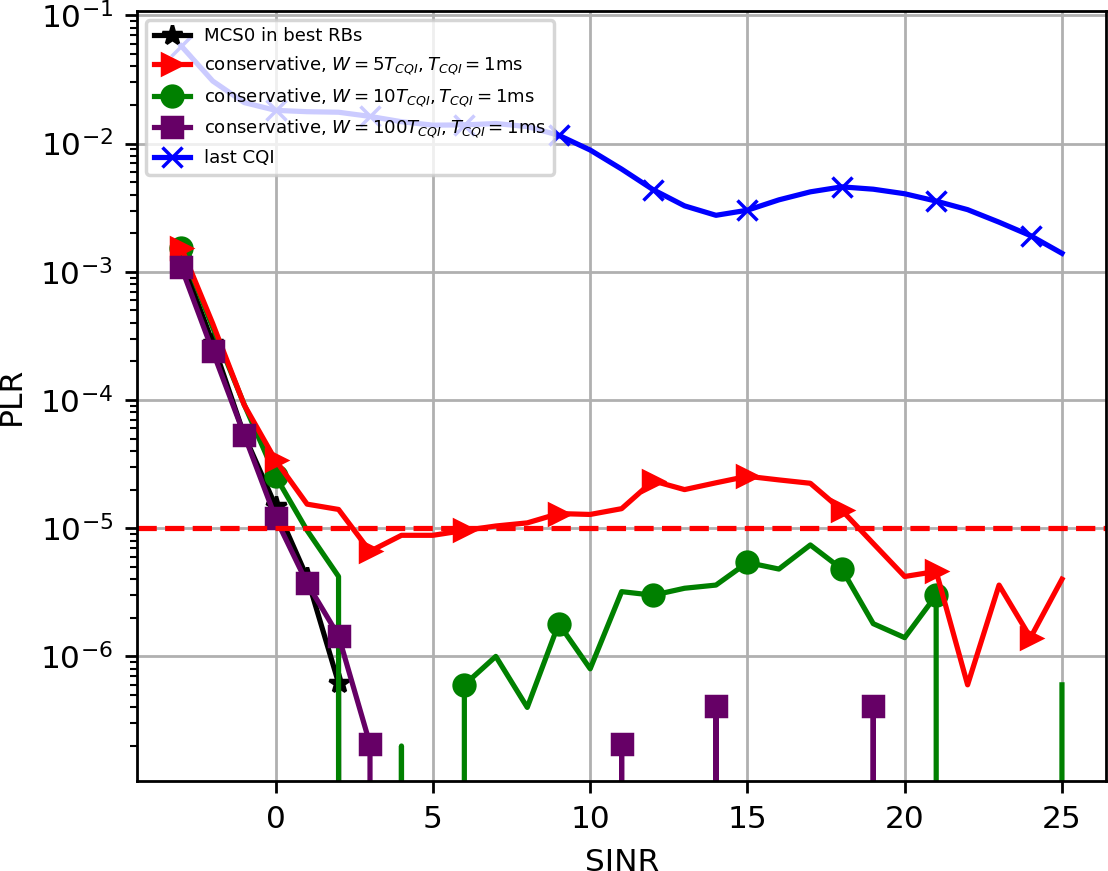}}
	\vfill
		\center{\includegraphics[width=0.7\linewidth]{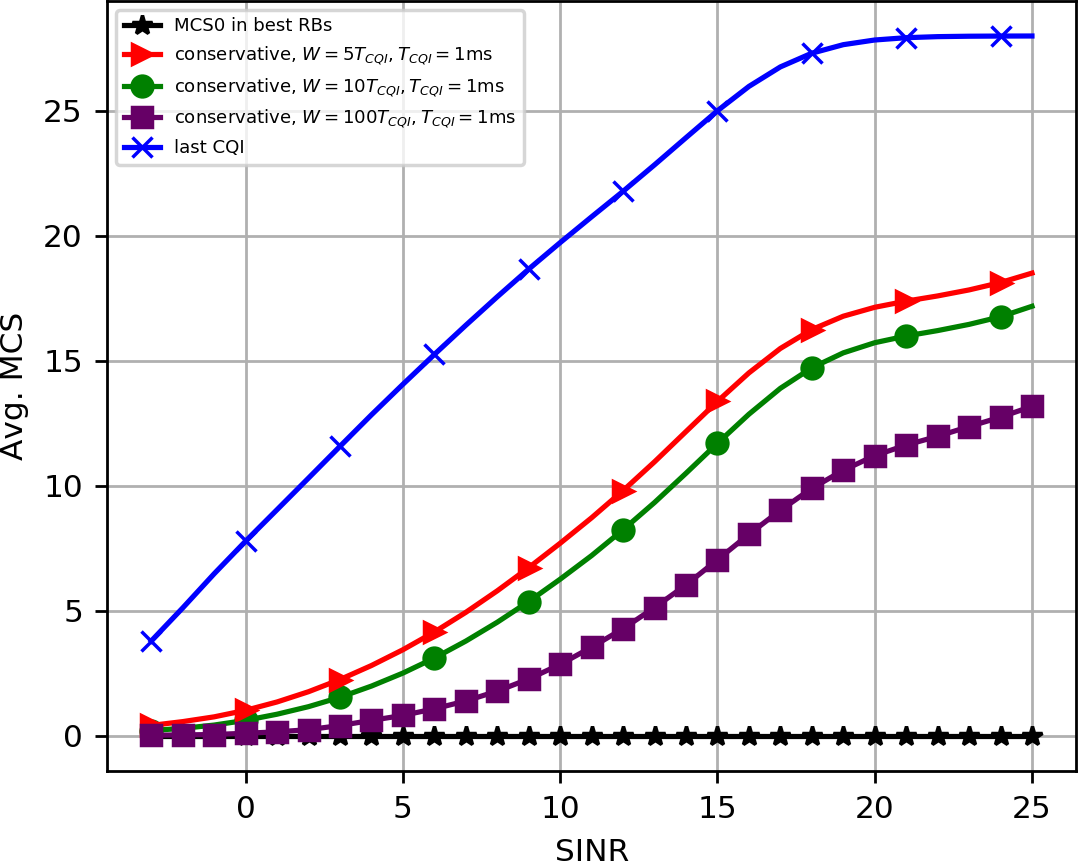}}
	\vfill
		\center{\includegraphics[width=0.7\linewidth]{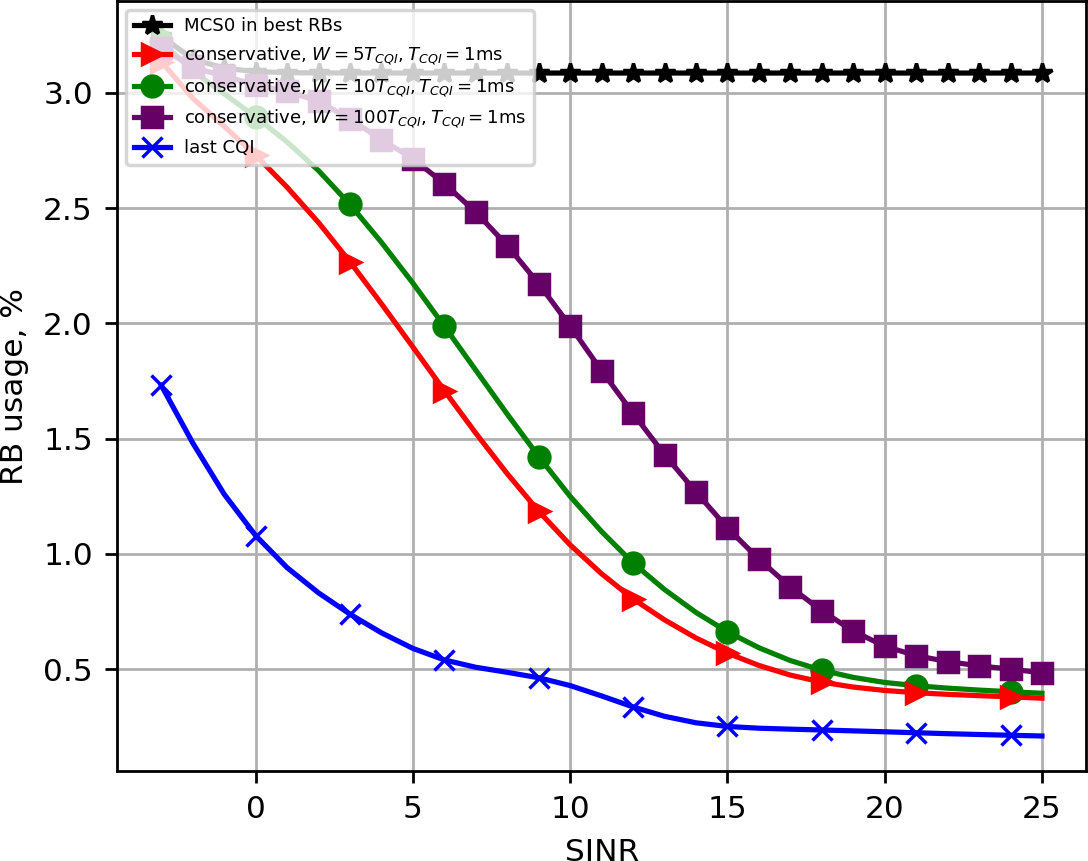}}
	\vspace{-0.3cm}
	\caption{Influence of WND for vehicle UE (60 kmph).}
	\label{fig:wnd_variation_EVA}
	\vspace{-0.5cm}
\end{figure}

\begin{figure}[!t]
		\center{\includegraphics[width=0.7\linewidth]{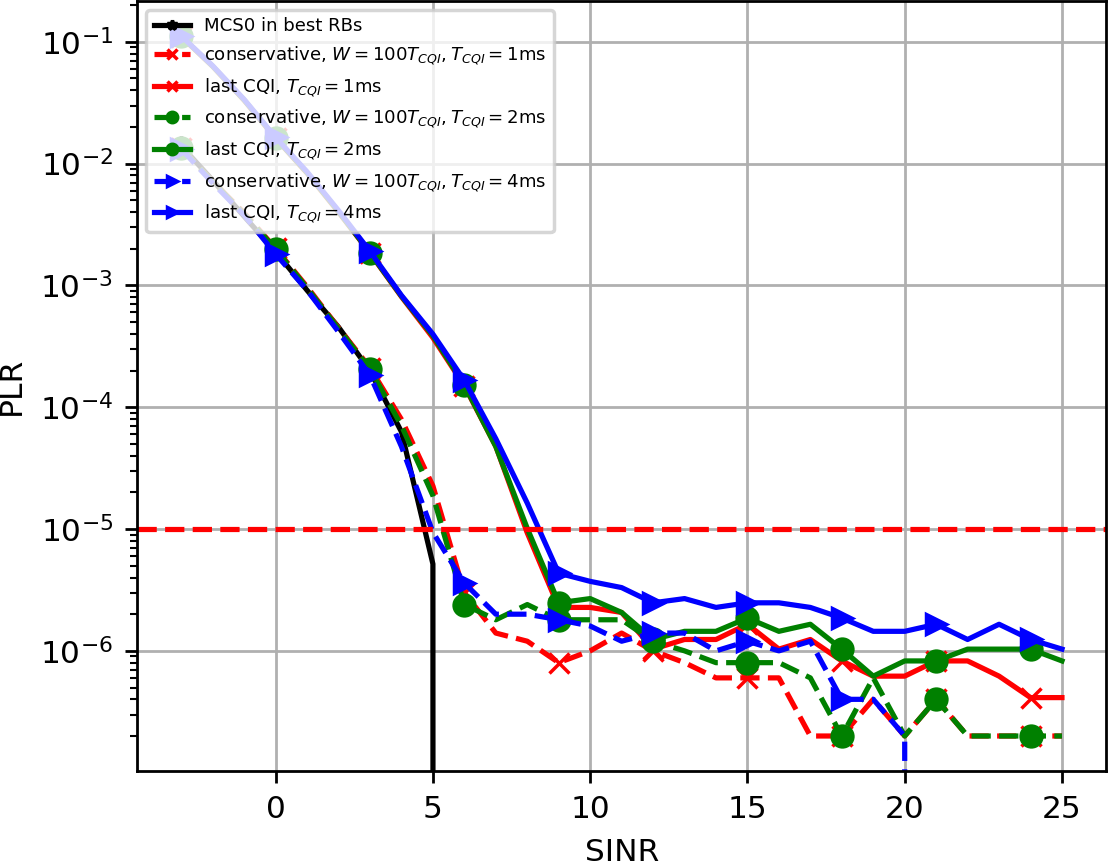}}
	\vfill
		\center{\includegraphics[width=0.7\linewidth]{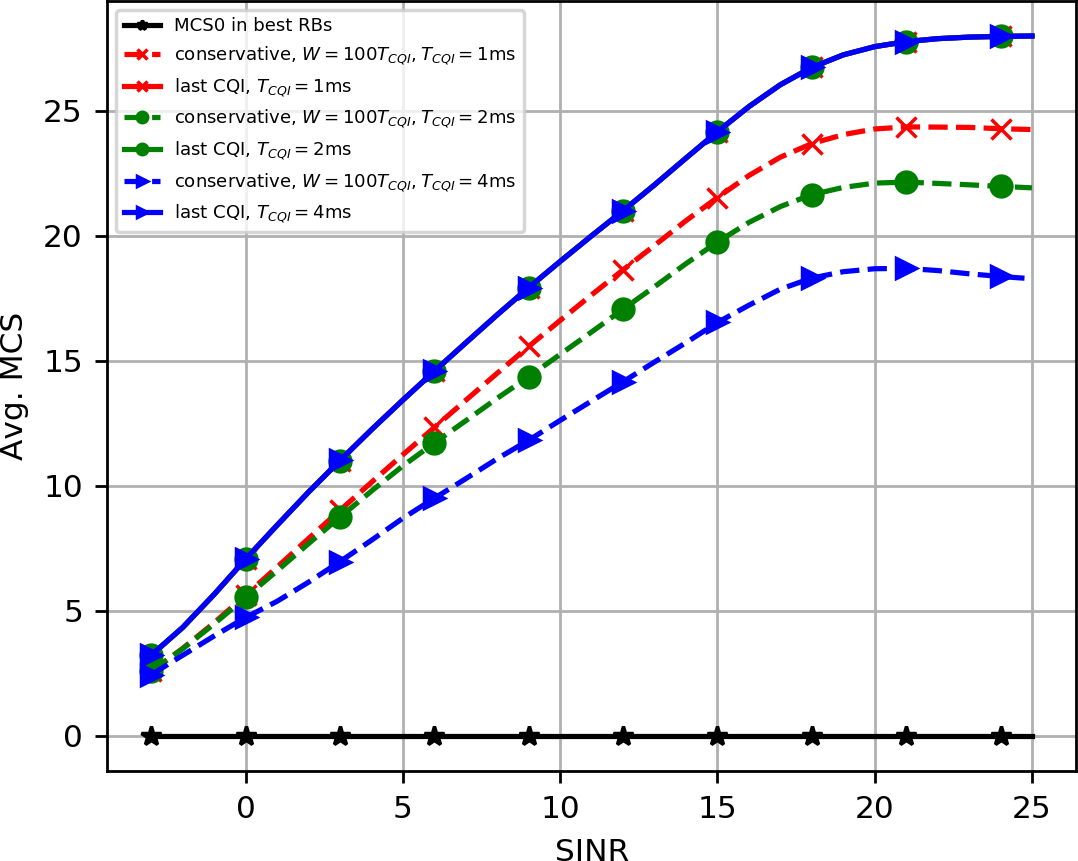}}
	\vfill
		\center{\includegraphics[width=0.7\linewidth]{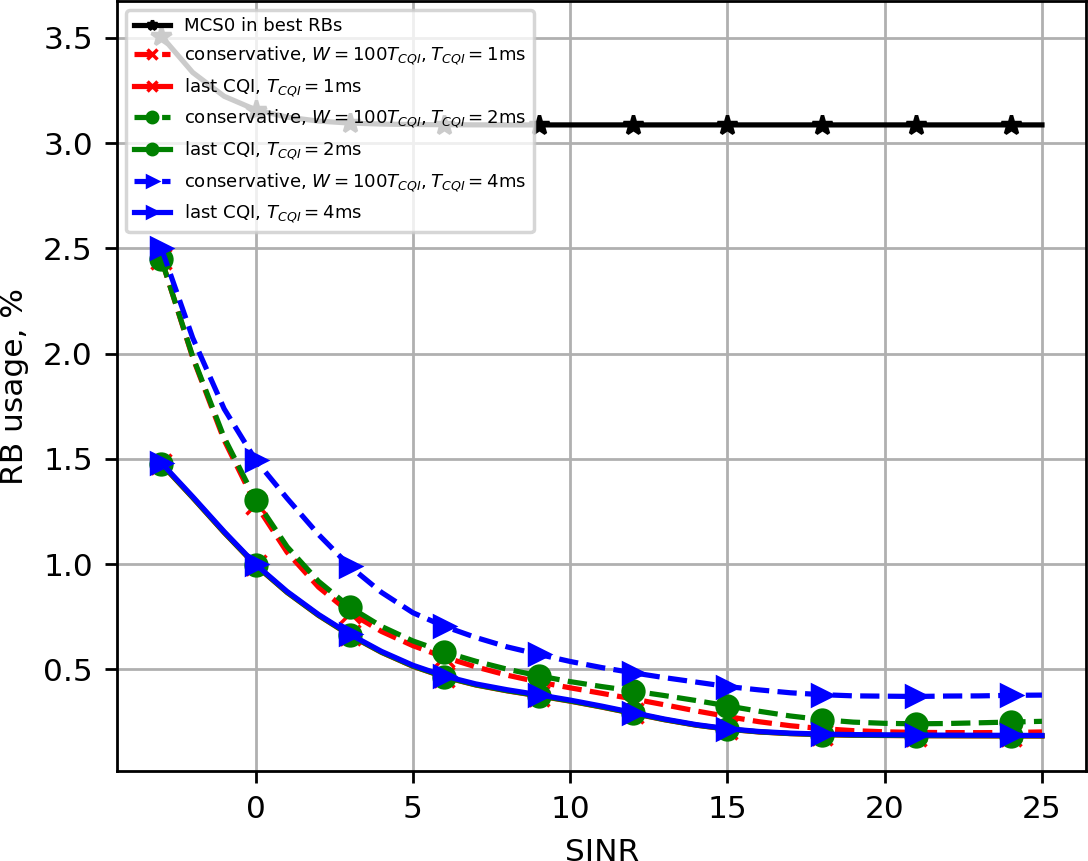}}
	\vspace{-0.3cm}
	\caption{Influence of CQI period for pedestrian UE (3 kmph).}
	\label{fig:tcqi_variation_EPA}
	\vspace{-0.5cm}
\end{figure}

\begin{figure}[!t]
		\center{\includegraphics[width=0.7\linewidth]{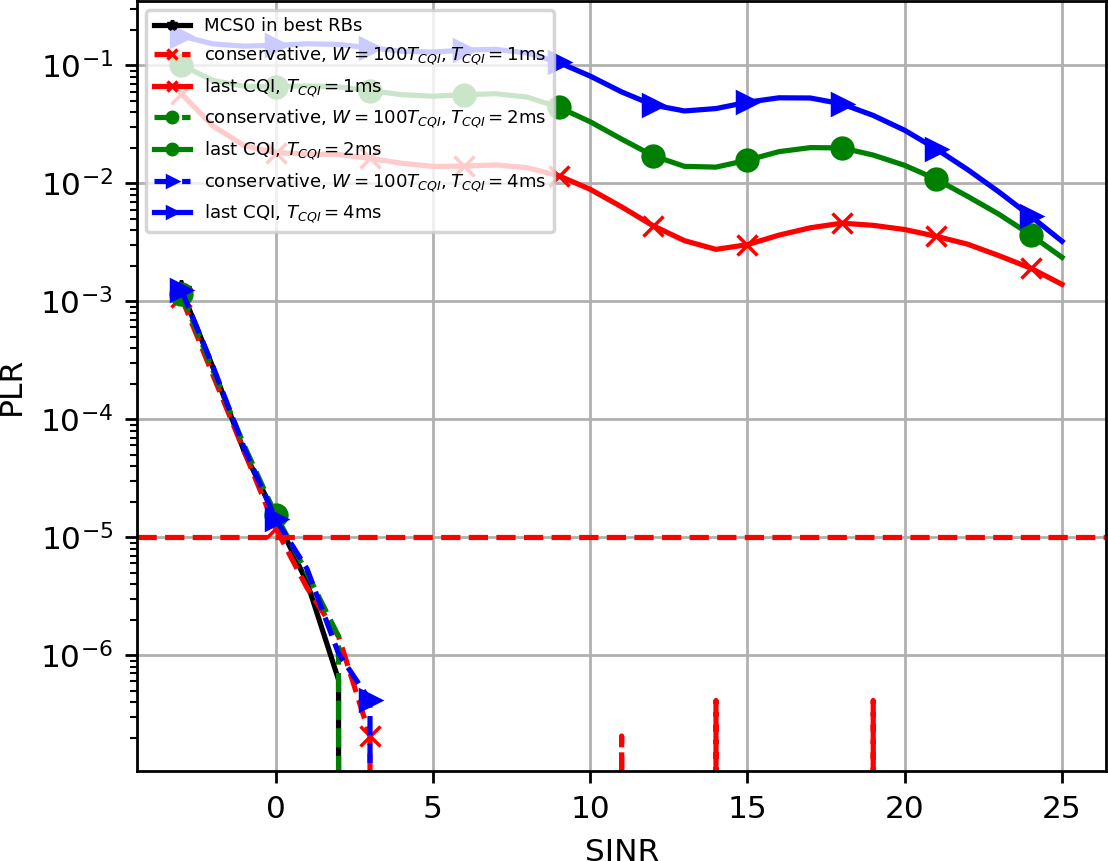}}
	\vfill
		\center{\includegraphics[width=0.7\linewidth]{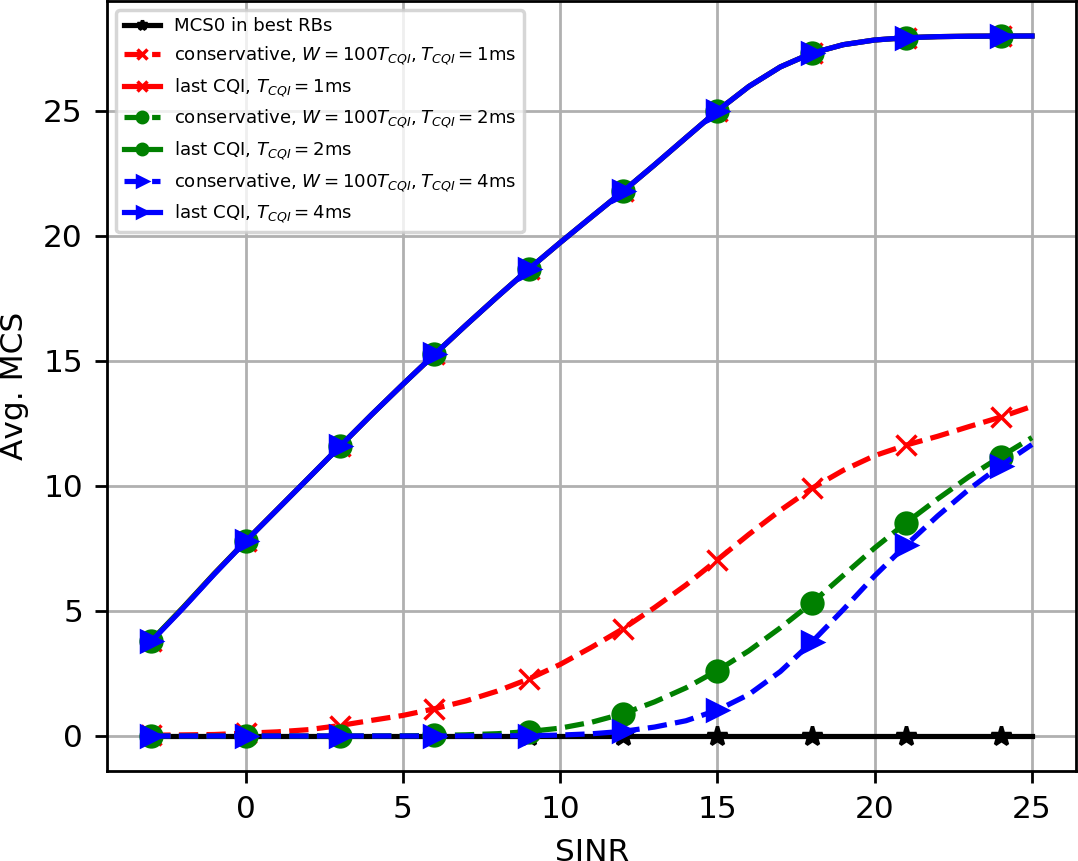}}
	\vfill
		\center{\includegraphics[width=0.7\linewidth]{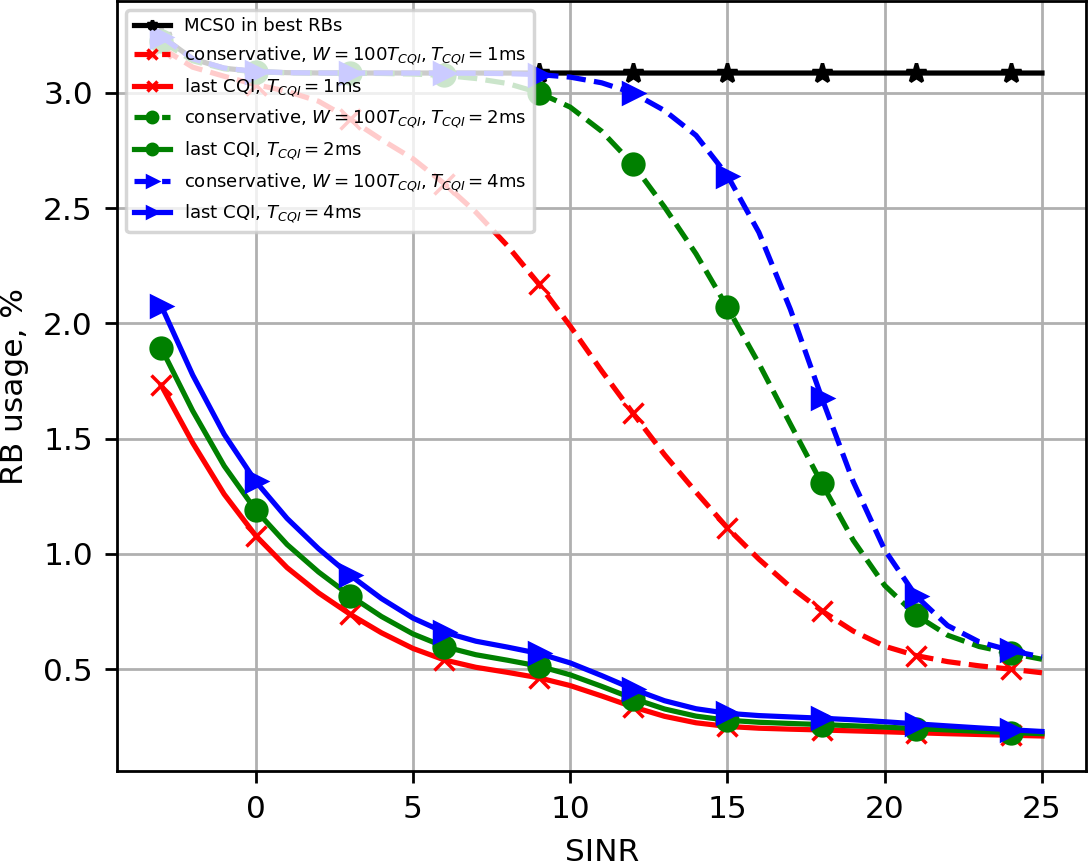}}
	\vspace{-0.3cm}
	\caption{Influence of CQI period for vehicle UE (60 kmph).}
	\label{fig:tcqi_variation_EVA}
	\vspace{-0.5cm}
\end{figure}

First, let us evaluate the influence of parameter $W$ on the proposed conservative algorithm performance. Fig.~\ref{fig:wnd_variation_EVA} shows results for a highly variable channel with Rayleigh fading corresponding to vehicles speeds of 60 kmph. We can see, that the channel estimation based on the last CQI provides very high PLR (higher than $10^{-3}$) even for a UE close to the gNB (with high geometry factor) because the gNB selects MCS values based on outdated CQIs. Although this solution consumes much less channel resource than other solutions, such a high PLR makes URLLC service impossible. 
In contrast, always selecting MCS 0 significantly reduces PLR but consumes much more channel resorce. Since MCS 0 provides the highest reliability for transmission, we can consider black curves as the lower bound of PLR which can be achieved with any solution. 

Let us consider the proposed conservative algorithm. We can see that its performance depends on $W$: the higher $W$ is selected, the lower PLR is achieved and the higher channel resource consumption is observed. These results can be explained by the fact that with a higher window the behavior of the algorithm becomes more conservative and it selects more robust MCS considering higher channel degradation. Figures show that small window ($5-10\:T_{CQI}$) is not enough to guarantee PLR $10^{-5}$. Hence, in the experiments below, we use $W = 100\: T_{CQI}$.  

Now let us evaluate the influence of the CQI period on the performance of various link adaptation algorithms. Figs.~\ref{fig:tcqi_variation_EPA} and~\ref{fig:tcqi_variation_EVA} show the results for fading corresponding to pedestrians with 3 kmph speed and vehicles with 60 kmph speed, respectively. The average selected MCS for the algorithm which only considers the last CQI does not depend on the CQI period since the average reported CQI remains the same with different CQI periods. Hence, the CQI period has almost no impact on the channel resource consumption for this algorithm. However, we can see a noticeable difference in PLR which increases with the CQI period since the reports become outdated for a higher CQI period. In contrast, for the proposed conservative algorithm, there is almost no difference in PLR for various values of the CQI period, but the channel consumption significantly increases with CQI period. This happens because $\Delta CQI$ in equation~\eqref{eq:cqi_diff} increases with the CQI period and hence the algorithm selects more robust MCS. Comparing with the solution which always selects MCS 0, we can see that the proposed algorithm significantly reduces channel consumption. In particular, for the pedestrian case, the reduction is observed even for cell edge UEs (i.e., with a low geometry factor). For high geometry factor, the difference in RB usage reaches six times for both pedestrian and vehicle cases. So, we can conclude that the proposed link adaptation algorithm allows achieving the same coverage as MCS 0 solution (i.e., PLR requirement is satisfied for UEs having low geometry factor) while significantly reducing channel resource consumption. 

\vspace{-0.2cm}
\section{Conclusion}
\label{sec:conclusion}

In this paper, we have proposed the novel conservative link adaptation algorithm for URLLC that estimates the worst-case channel degradation experienced by UEs. For that, the gNB keeps statistics of received CQI reports and calculates the maximum channel quality degradation over a time window. We compare this algorithm with two reference solutions: (i) to select MCS based on the latest received CQI report, and (ii) to select always the most robust MCS (i.e., MCS 0). The results obtained with NS-3 simulator show that the first reference solution does not allow the gNB to satisfy strict URLLC requirements, especially in highly mobile scenarios, while the second one consumes too much channel resource. In contrast, the proposed approach allows satisfying URLLC requirements for a wide range of geometry factor values (i.e. provide high network coverage) and at the same time provides up to 6 times reduction in channel consumption compared to the second reference solution. 

\bibliographystyle{IEEEtran}
\bibliography{main}

\end{document}